\documentclass[a4paper,11pt]{article}
\pdfoutput=1
\usepackage{jcappub}
\usepackage[T1]{fontenc}
\usepackage{subfig}
\usepackage{caption}
\usepackage{hyperref}
\usepackage{amsmath}
\usepackage{mathabx}
\usepackage{mathtools}
\usepackage{tablefootnote}
\usepackage{soul}
\usepackage{ulem}
\usepackage{color}
\usepackage{float}
\usepackage{makecell}
\usepackage[usenames,dvipsnames,svgnames,table]{xcolor}
\usepackage{indentfirst}
\usepackage{leftidx}

\title{Giants eating giants: Mass loss and giant planets modifying the luminosity of the Tip of the Giant Branch}

\author[a,b]{Raul Jimenez,}
\emailAdd{raul.jimenez@icc.ub.edu}

\author[c]{Uffe Gr{\aa}e J{\o}rgensen,}
\emailAdd{uffegj@nbi.dk}

\author[a,b]{Licia Verde}
\emailAdd{liciaverde@icc.ub.edu}

\affiliation[a]{ICC, University of Barcelona, Marti  i Franques, 1, E-08028 Barcelona, Spain.}
\affiliation[b]{ICREA, Pg. Lluis Companys 23, Barcelona, E-08010, Spain.}
\affiliation[c]{Niels Bohr Institute \& Centre for Star and Planet Formation, University of Copenhagen, {\O}ster Voldgade 5, 1350 Copenhagen, Denmark}

\abstract{During the red giant phase, stars loose mass at the highest rate since birth. The mass-loss rate is not fixed, but varies from star-to-star by up to 5\%, resulting in variations of the star's luminosity at the tip of the red giant branch (TRGB).  Also, most stars, during this phase, engulf part of their planetary system, including their gas giant planets and possibly brown dwarfs. Gas giant planet masses range between 0.1 to 2\% of the host star mass. The engulfing of their gas giants planets can modify their luminosity at the TRGB, i.e. the point at which the He-core degeneracy is removed. We show that the increase in mass of the star by the engulfing of the gas giant planets only modifies the luminosity of a star at the TRGB by less than 0.1\%, while metallicity can modify the luminosity of a star at the TRGB by up to 0.5\%. However, the increase in turbulence of the convective envelope of the star, i.e., modification of the mixing length, has a more dramatic effect, on the star's luminosity, which we estimate could be as large as 5\%. The effect is always in the direction to increase the turbulence and thus the mixing length which turns into a systematic decrease of the luminosity of the star at the TRGB. We find that the star-to-star variation of the mass-loss rate will dominate the variations in the luminosity of the TRGB with a contribution at the 5\% level. If the star-to-star variation is driven by environmental effects --as it is reasonable to assume--,   the same effects can potentially create  an environmentally-driven  mean effect on the luminosity of the tip of the red giant branch of a galaxy. Engulfment of a brown dwarf will have a more dramatic effect. Finally, we touch upon how to infer the frequency, and identify the engulfment, of exoplanets in low-metallicity RGB stars through high resolution spectroscopy as well as how to quantify mass loss rate distributions from the morphology of the horizontal branch.}
\begin{document}
\maketitle

\section{Introduction}\label{sec:intro}

As stars with masses between 0.8 and 2 M$_{\odot}$ turn off the main sequence, they start to burn hydrogen in the shell  around a degenerate core of helium supported by electron degeneracy pressure. For a fixed chemical composition, turbulence in the convective envelope and mass, the electron degeneracy is lifted at a fixed core mass via rapid triple$-\alpha$ nuclear burning into carbon and oxygen. Because there is a very well-defined relationship between core mass and luminosity (\cite{Kippenhahn,Marigo}), this results in a very well-defined luminosity of the tip of the red giant branch (TRGB) very evident in a  Color-Magnitude diagram. Once the degeneracy is lifted, the stellar luminosity decreases dramatically as helium burning starts in the core and the star settles on the horizontal branch~\cite{Kippenhahn}. This fact has been exploited  as a  calibration to measure astronomical distances; recently, this technique has been used to obtain $2-3$\% precision in the determination of the Hubble constant~\cite{Freedman}. Because of the critical role that the distance determination from the TRGB plays in the current debate of the local value of $H_0$ (see e.g. Ref.~\cite{VerdeH0}), we investigate possible effects of other environmental parameters  besides the intrinsic properties of the host star itself.

It has been known for a while that mass loss at the TRGB will produce a spread in its  luminosity at the 5\% level (see Fig. 2 and Table~1 in \cite{Jimenez95}). This spread is needed to reproduce the morphology of the horizontal branch~\cite{Jimenez96,Jimenez2004} for a large range of metallicities (see Fig.~3 in Ref.~\cite{Jimenez2004}). Ref.~\cite{Jimenez96,Jimenez2004} exploited this fact to predict a distribution function in the mass-loss efficiency of red giants (expressed as a star-to-star variation in Reimer's~\cite{Reimers} parameter $\eta$), based on fits to the horizontal branch morphology~\cite{Uffe93,Jimenez95}.  Besides this, one effect that has not been considered to-date in this context is the engulfment of potential gas giant planet(s) in the red giant star planetary system. 

This can happen if the star reaches a radius that encompasses the orbit of the planets.
The authors of Ref.~\cite{Sackmann} found that the Sun itself will reach a radius of 0.8\,AU at the TRGB, which, for the general population of $\sim 1 M_{\odot}$ stars now  climbing up the main sequence, would be large enough to enclose 30\% of all the known radial velocity giant exoplanets listed in {\it exoplanet.eu}. This is further supported by the work of Ref.~\cite{Uffe91} who reviewed the results of a number of different general stellar evolutionary grids, and concluded that, by interpolating their results to the solar value, most grids predicted a solar radius at the TRGB to be  only slightly below 1\,AU, just as Ref.~\cite{Sackmann}'s direct computation, and with the highest estimate being 1.02\,AU. However, Ref.~\cite{VLiviob} found that even planets in considerably larger orbits than the size of the host star itself  will be dragged into the star due to tidal interaction between the star and the planet, and eventually will be engulfed by the star. The authors of Ref.~\cite{VLiviob} found that this engulfment is more efficient for more massive planets and less massive stars, and that the planets will be engulfed well before the star reaches the TRGB (see Fig.2 in Ref.~~\cite{VLiviob}). Ref.~\cite{VLiviob} estimates that planets in orbits 
as large as 3 AU will be engulfed into the host star because of tidal interactions\footnote{This is a resonnable estimate; they  assumed  that a solar mass star reaches  a radius of 1.1\,AU at the TRGB and  used a Reimer's type mass loss with $\eta$=0.6. The maximum radius adopted is probably a slight overestimate, and the  adopted $\eta$ value is slightly too large to fit the globular cluster morphology;  a lower more realistic value of $\eta$ $\approx$ 0.4 would imply that exoplanets with a slightly larger orbital radius would be engulfed too. Hence this  is qualitatively compensating for by the slightly too large stellar radius adopted.}.

The repository at {\it exoplanet.eu} lists that 85\% of all radial velocity discovered giants are within 3\,AU. Higher-mass stars reach a smaller TRGB stellar radius than 1\,M$_{\odot}$ stars and would therefore engulf a smaller percentage of gas giant planets, but for any stellar initial mass function, low mass stars at 1\,M$_{\odot}$ dominate in numbers. Since of the order of 10\% of all stars searched with the radial velocity technique show one or more giant exoplanets, we should expect that the distribution of TRGB luminosities include of the order of $8.5\sim$10\% of stars affected by engulfment of their  gas giant exoplanets.

There are several effects of a star engulfing its gas giants. 

The most obvious one is an increase in the mass of the star: this will be in the range 0.1-2\% (for a Jupiter mass planet orbiting a Sun-like star to a 10 Jupiter mass exoplanet orbiting an M dwarf). 

The second effect is an increase in the metallicity of the stars. For the engulfment of Jupiter itself, a reasonable estimate of the value of heavier elements would be $10$ M$_{\Earth}$ metal/rock/water-ice (as listed in Ref.~\cite{LoddersFegley}) in the core plus an enhancement of a factor 3 in Z as compared to $Z_\odot$ in the envelope of Jupiter  (estimated from the atmospheric abundance of CH$_4$), which would add up to approximately $20$ Earth masses---and a minimum of $10$  Earth masses---of heavy elements in Jupiter. Stars probably have up to $5 M_{\Earth}$ of rocky planets in orbits of 1 to 2 AU radii, which,  according to the above  argument, would lead to  the addition of $\sim$5 to 25 M$_{\oplus}$ of heavy material to the star during the red giant branch. For Sun-like metallicity, this  corresponds to a relative increase in the value of Z of the host stars at the TRGB of a few tenths of a percent compared to their main sequence value (5\,M$_{\oplus}$ $\sim$\,0.1\% and 25\,M$_{\oplus}$ $\sim$\,1\%--recall that the Sun metallicity is just 1.5\% of its mass)  For our fiducial stars in the figures below ($Z$ = 0.003) it would correspond to a few \%, and for a typical star in the Large and Small Magellan clouds ($Z$=0.0007 and $Z$=0.0004, respectively; \cite{Haschke2012}) engulfing 5 to 25 M$_\oplus$ solid material would increase the TRGB stellar metallicity
 of the order of 10\%, assuming total mixing, making it easily visible in the spectra of the RGB stars. 

The third effect is an increase in the turbulence of the convective layer of the star\footnote{For the star mass of interest ($0.8-2.0$ M$_{\odot}$) stars always have a radiative core and a convective envelope, with the relative mass of the envelope decreasing with increasing total stellar mass} as the planet moves at supersonic speed with respect the velocity of the turbulent outer layers of the red giant.

In this short note, we evaluate quantitatively these environmental effects, mass-loss variation and gas giant engulfment, in order to understand how they affect the star's  luminosity at the TRGB. Calculations can be done quantitatively for an individual star. The mean effect on the population of stars defining the  observed TRGB in a Color-Magnitude diagram, the possible  galaxy-to galaxy variation  and how it propagates  into the distance estimate at this time can be discussed only qualitatively. We highlight the importance of modeling these environmental effects in detail to understand the robustness of the TRGB as a distance indicator with, or better than, 1-\% accuracy. We leave a quantitative calculation to future work.

\section{The theoretical tip of the red giant branch}

The tip of the red giant branch (TRGB) is determined by the removal of the electron degeneracy of the He core during H shell burning. H burning on the shell causes the envelope of the star to expand and become a giant. Besides the total mass of the star, other parameters affect the tip of the red giant branch. In what follows, we use the accurate fitting formulas presented  in Ref.~\cite{JimenezMacdonald96} from an extensive grid of stellar evolutionary tracks computed with the JMSTAR stellar code. The luminosity of a start at the TRGB is 
\begin{eqnarray}
\log_{10} (L_{\rm TRGB}/L_{\odot}) & = & 2.837 - 0.071M + 0.045 M^2 +0.0580 \alpha - 0.059 \alpha M \nonumber \\ 
& &  - 0.182 \alpha^2 + 45.163 Z + 0.581 \alpha Z - 6435.45 Z^2 
\label{eq:lum}
\end{eqnarray}
where $L_{\rm TRGB}$ is the luminosity of the star at the TRGB, $M$ is the mass of the star in $M_{\odot}$, $\alpha$ is the mixing-length parameter in 1D convection theory and finally $Z$ is the fractional percentage of elements heavier than H and He in the star. The accuracy of this formula is 0.0007 for $\log (L_{\rm TRGB}/L_{\odot})$\footnote{The stellar models are accurate at better than sub-\% precision to estimate the differential variations of the luminosity of the TRGB as a function of input parameters.}.

It is also useful to compute the time to helium flash in Gyr; this is given by
\begin{equation}
\log_{10} t = -0.207 - 3.691 \log M + 11.327 \log (1.76-Z) + 0.870 \log (0.0086+Z).
\label{eq:time}
\end{equation}
This formula has a $5\%$ uncertainty. In both cases we have fixed the primordial Helium fraction to $Y=0.24$. \footnote{Variations in the value of  $Y$ will also affect the luminosity of the TRGB but we will not explore them here. The primordial helium fraction here is not necessarily  identical to the cosmological one. If PopIII stars could have  enriched the medium with Helium, it is  the resulting Helium fraction that matters here. Hence in principle  there could be galaxy-by-galaxy variations of $Y$.}

\begin{figure}
\centering
\includegraphics[width=0.45\columnwidth]{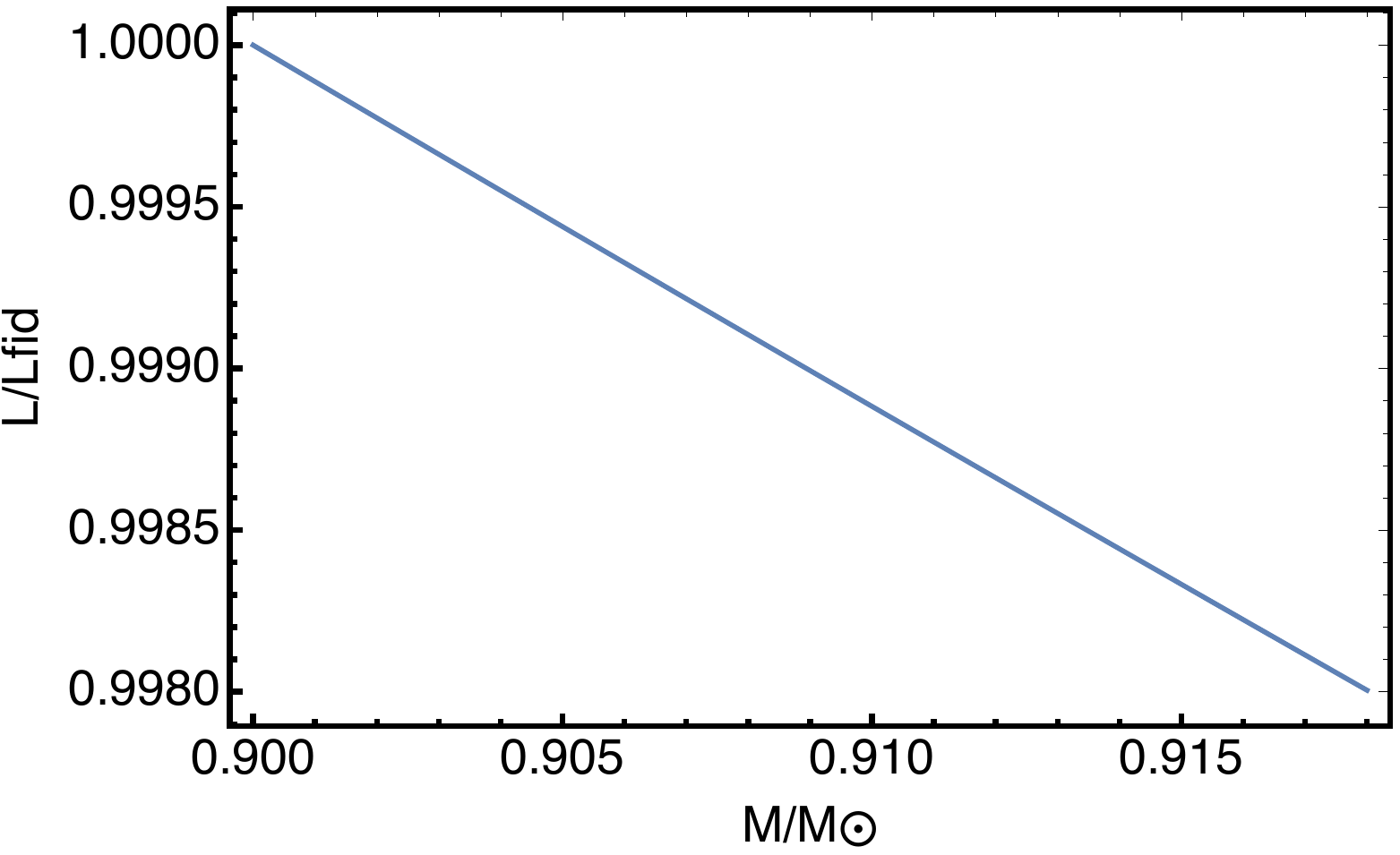}
\includegraphics[width=0.45\columnwidth]{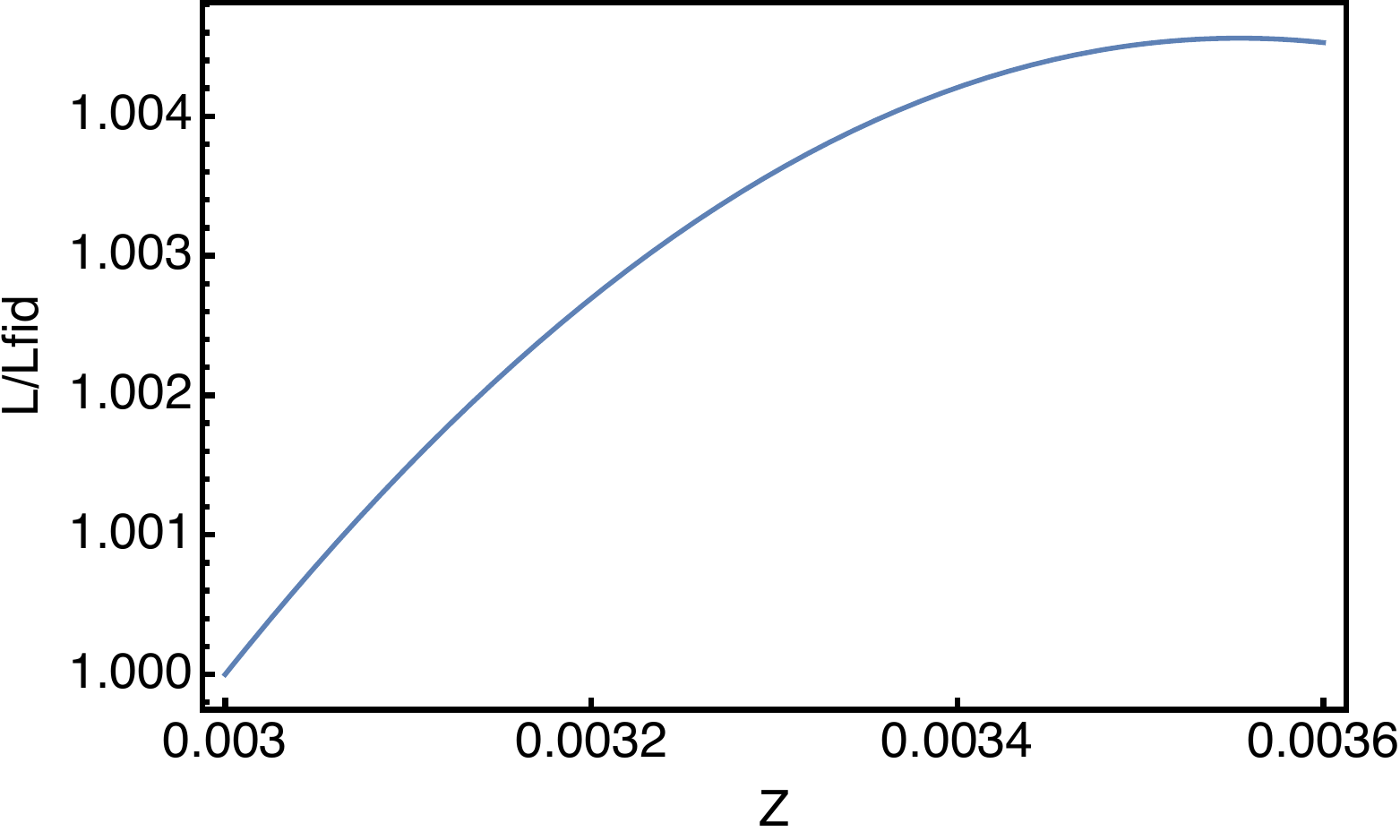}
\includegraphics[width=0.45\columnwidth]{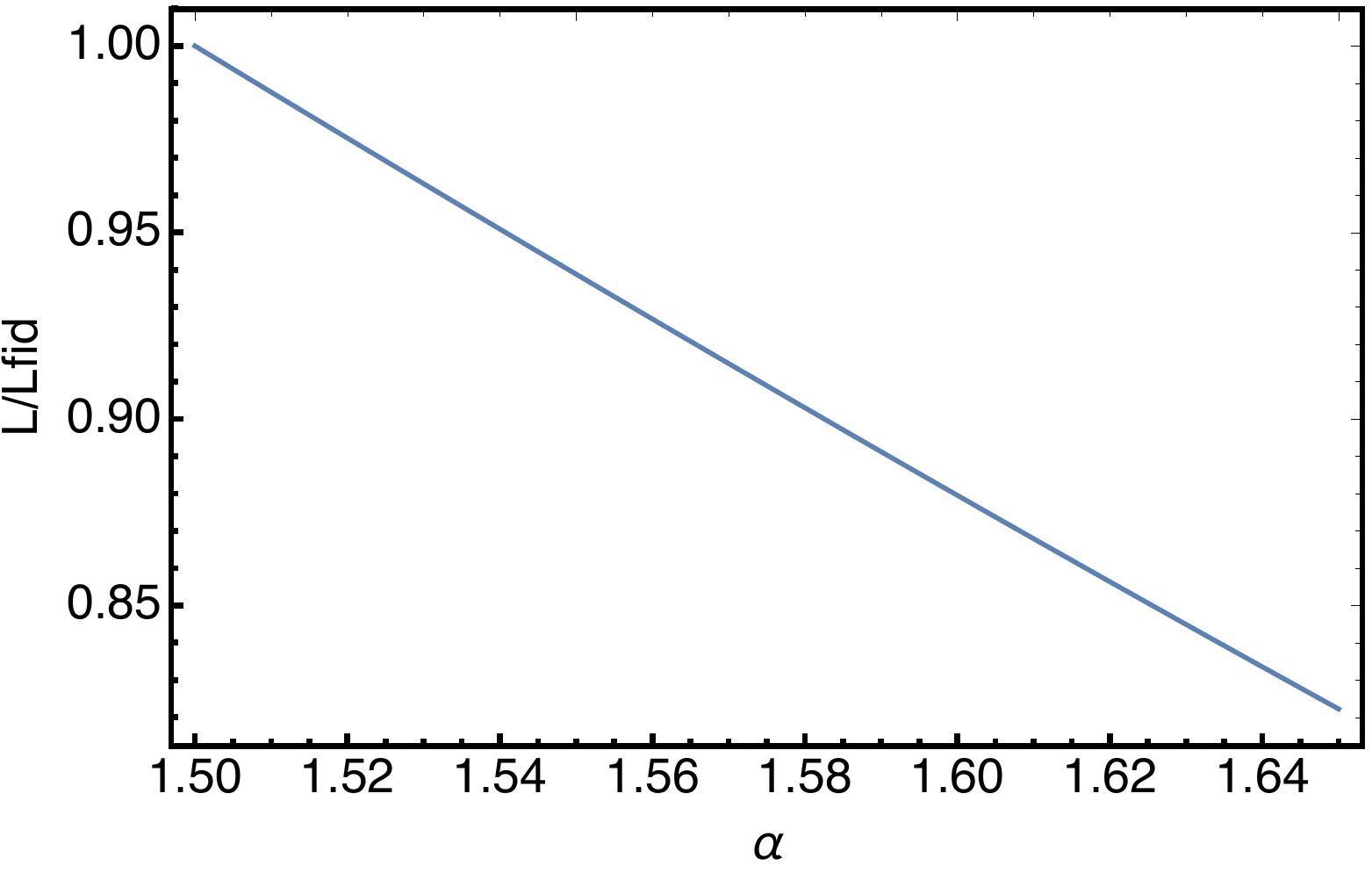}
\includegraphics[width=0.45\columnwidth]{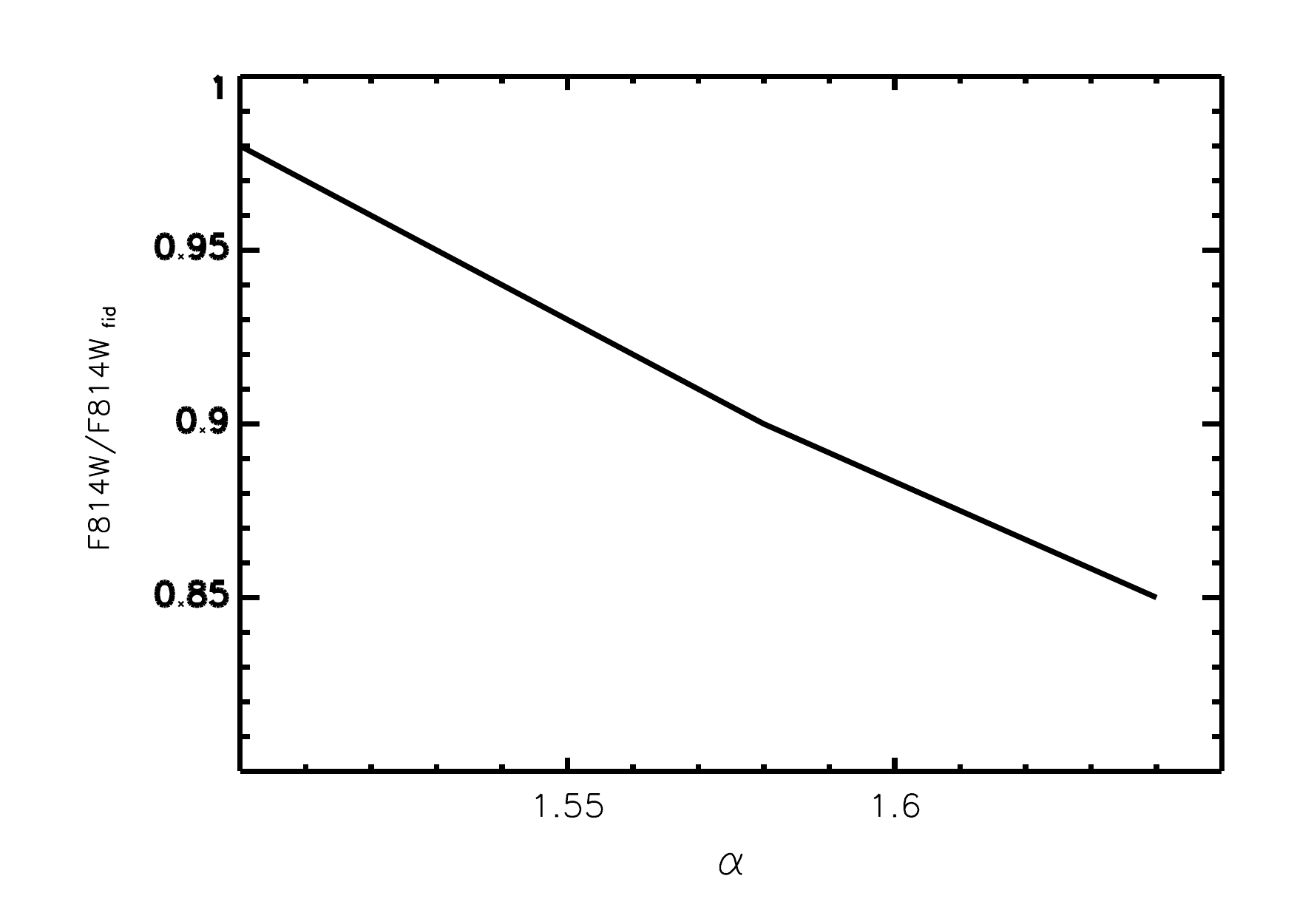}
\caption{Fractional change of the luminosity of a star at the TRGB as a function of changes in the total mass of the star (upper-left panel), metallicity (upper-right panel) and mixing length (bottom-left  panel), measured relative the luminosity, $L_{\rm fid}$, of our fiducial TRGB star with (M/M$_\odot$, Z, $\alpha$, $\eta$) = (0.9, 0.003, 1.5, 0.35). The bottom-right panel shows the relative variation in the WF3 HST IR filter $F814W$ which is typically used to determine the TRGB observationally.}
\label{fig:models}
\end{figure}

First note that the time to the Helium flash only depends on the mass and metallicity of the star. On the other hand, the luminosity of a star at the TRGB does have an additional dependence on the mixing-length parameter.

Using eq.~\ref{eq:lum} we computed the variation in luminosity when the mass ($M$), metallicity ($Z$) or mixing length ($\alpha$) change by an amount between 0 and 20\%. Results can be seen in Fig.~\ref{fig:models}.  Roughly $d\ln L/d\ln M\sim 0.13$ hence a 1\% change in mass will correspond to $\sim 0.1\%$ change in luminosity; $d\ln L/ d\ln Z$ seems to show two regimes, at $Z<0.0032$ we have $d\ln L/ d\ln Z\sim 0.032$ hence a 10\% change in mentality, Z, will corresponds to a $\sim 0.3\%$ change in luminosity, but at $Z>0.0032$ the luminosity is increasingly less sensitive to  changes in $Z$. Thus we expect that  changes of the luminosity of  the TRGB  due to variations in the mass  of the star because of the engulfment of the gas giants will translate into changes of less than 0.1\%, so truly negligible. Changes due to the expected increase in metallicity due to the engulfment of stars can, on the other hand, be as much as 0.5\%, which could be of relevance in the era of 1\% accuracy in astronomical distances but it is not a worry at present time. On the other hand, the luminosity of a star at the TRGB depends strongly on the mixing length parameter, $d\ln L/d\ln \alpha \sim 0.75$. In Sec.\ref{sec:engulfing-mixing} we quantify the effect of engulfing the gas giant on the mixing length parameter and thus in the  star's luminosity at the TRGB.

\section{Variations in the mass-loss rate at the TRGB.}

\begin{figure}
\centering
\includegraphics[width=0.6\columnwidth]{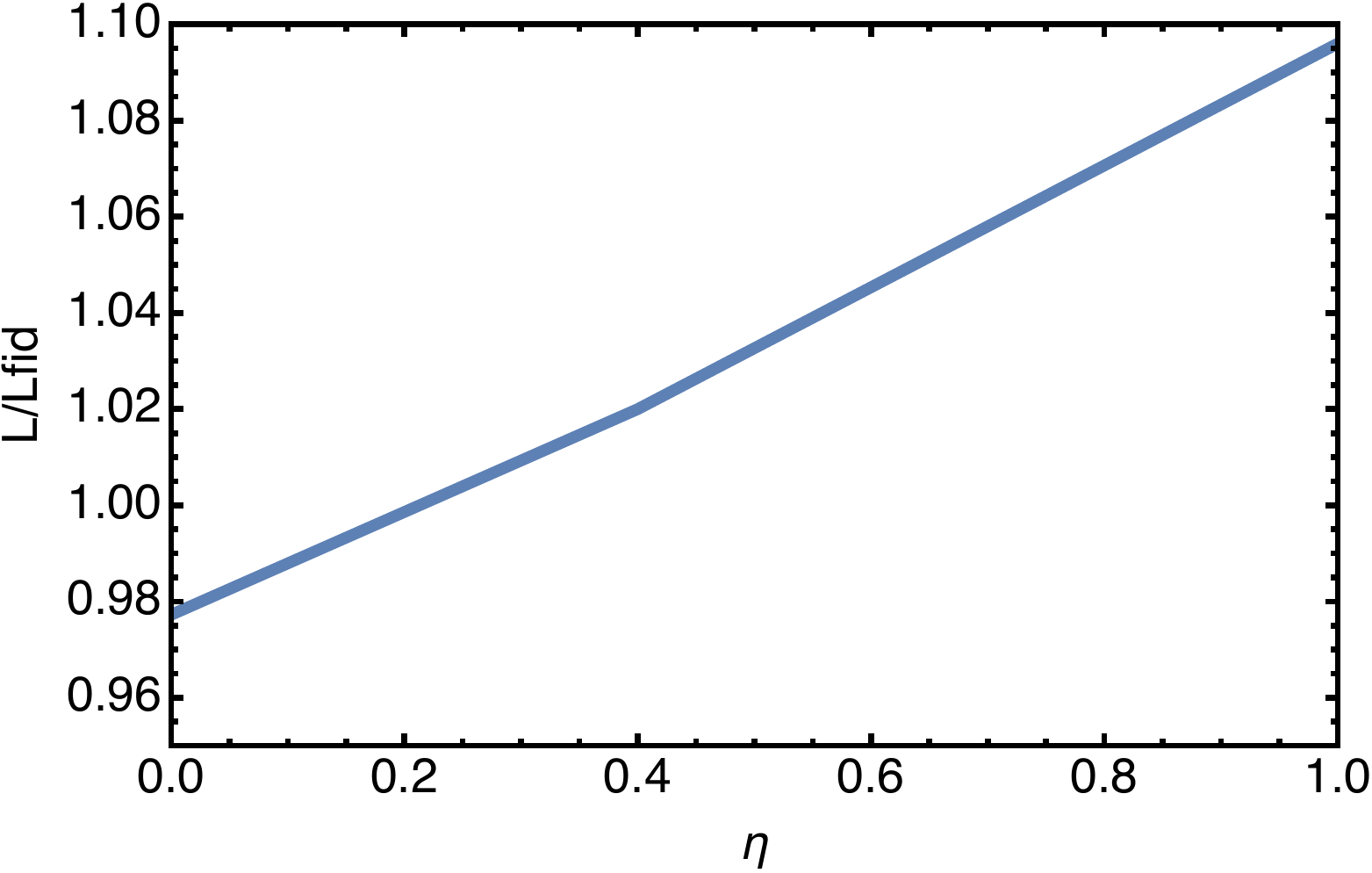}
\caption{Effect on the luminosity of the TRGB of variations on the $\eta$ parameter ($0-1.0$) in the Reimer's mass-loss formula.}
\label{fig:massloss}
\end{figure}

The mass loss at the tip of the giant branch takes place at a high-rate due to radiation pressure is an environmental parameter. The models of Ref.~\cite{Hofner,Hofner2}  show that dynamical mass loss during pulsation is due to changes in the wind driven mass loss as a function of the changing effective temperature during the pulsation and associated changes in the dust and molecular opacities that drive the wind. In this case,   dynamical mass loss, is also dust driven and thus metallicity dependent. A recent review on mass loss processes in stars can be found in Ref.~\cite{Willson}.

An empirical formula to describe this process is Reimer's mass loss~\cite{Reimers}
\begin{equation}
\dot M = -4 \times 10^{-13} \eta \frac{L/L_{\odot}}{(g/g_{\odot}) (R/R_{\odot})} \frac{M_{\odot}}{\rm yr},
\end{equation}
where $L, g, R$ are the luminosity, surface gravity and star radius respectively and $\eta$ is the (unkown) mass-loss rate parameter.
 Note that mass-loss rate grows linearly with luminosity, which is highest ($\sim 1000$ L$_{\odot}$) at the TRGB.  In Refs. \cite{Uffe93,Jimenez96} it was shown that, in order to describe the (extended) morphology of the horizontal branch (its spread in temperature), there must be   star-to-star variations in massloss. These variations translate in a range of $0 < \eta < 1$ with a broad distribution between these values. Using the detailed computations in Ref.~\cite{Jimenez95} we have evaluated the variation of a star's  luminosity at the TRGB as a function of $\eta$. Results are shown in Fig.~\ref{fig:massloss}. Note that in this case the effect is significant, of about 5\%. The spread of $\eta$ depends on the environmental conditions like metallicity and age (see Fig.~3 in Ref.~\cite{Jimenez2004}). Inducing a star-to-star variation of the luminosity by about 5\%. While the galaxy-to-galaxy variation is expected to be smaller because of the large number of stars involved, environmental conditions could still potentially affect the TRGB luminosity  via a modulation of $\eta$. If  the maximum modulation is achieved across different galaxies then the effect on the inferred distances could be as large as $\sim$ 2\%. The galaxy-to-galaxy modulation of  $\eta$ therefore should  be quantified in detail. If $\eta > 1$ then most of the envelope, for a $\sim$ solar mass star,  is removed and the star evolves directly into the white dwarf cooling sequence. This will result in a reduction of a factor $10$ of the luminosity of the TRGB; these values of $\eta$, while possible, are very rare.

\section{Engulfing of gas giants by giant stars and its impact on mixing length}
\label{sec:engulfing-mixing}

It is possible to estimate the effect that engulfing a giant planet will have on the convective envelope of a red giant star from analytical considerations\footnote{3D hydro-dynamical modeling of this process is extremely challenging as the convective envelope of the star is fully turbulent with Reynolds number $\sim 10^{10}$.}.

Let us start with the qualitative picture. A gas giant planet at a distance of about 0.5 AU (1 AU, 3 AU)  from its 1$M_{\odot}$ host star, will orbit it at about 40 km/s ( 30 km/s, $\sim$ 20 km/s).  On the other hand the gas giant outer envelope moves at 5 km/s. The velocity of the "blobs" in the outer convective layer of the atmosphere is of the same order ($\sim  5$ km/s).  The planet's velocity, $v_p$ is therefore a factor  few to $\sim 10$ greater than that of the stellar envelope. The mass of the envelop is $M_{env}\sim 0.2 M_{\rm star}$ and  if the mass of the planet is  $M_{\rm p}\sim 1\% M_{\rm star}$,  this yields $M_{\rm p}/M_{env}\sim  0.05$. Hence, the deposition of the planet's orbital energy $\sim 1/2 M_p v_p^2$, is a  significant disturbance to the stellar envelop.  As the giant planet is engulfed it will stir the convective envelope, like a spoon stiring a cup of tea. The  stirring effect  will last for one dynamical time of the envelop which is significantly larger than the time the star spends as a TRGB star. 

In Ref.~\cite{VLiviob} it is shown that planet engulfment is likely to happen  before the star on the red giant branch reaches its maximum radius and where the luminosity is about $100\, L_{\odot}$. This is also shown in the simulations by Ref.~\cite{Cantiello}: the most likely point of engulfment is at luminosity of about  $100\, L_{\odot}$, this is half way along the red giant branch. They also calculate that it will take about $10^4$ orbits before the planet loses most of its angular momentum and it sinks to the radiative core of the star.

Several effects are expected to happen: gas dynamic of the planet engulfment alters the star's evolution \cite{Soker84,Sandquist98,SiessLivio99a,SiessLivio99b,Staff16,Metzger17,Cantiello,Stephan}. The angular momentum deposited in the stellar envelop can  enhance significantly  stellar rotation \cite{Soker98,SiessLivio99b,Zhangpenev14,Privitera16a,Privitera16b,Stephan}. Stellar ejecta can produce transients\cite{SokerTylenda06,Metzger2012,Yamazaki2017,Stephan}.
Ref.~\cite{Cantiello} studies the effect  on the  Hertzsprung$-$Russell  (HR) diagram of planet engulfment by stars, by  comparing the orbital decay power to stellar luminosity. In particular they show that stars at $L=100 L_{\odot}$ receive a perturbation of order 100\% if they engulf a Jupiter-size planet.   

In Ref.~\cite{Stephan} possible observational signatures of planet "consumption" were considered for stars of different  masses and evolutionary phases.  In a red giant the planet will dissipate its orbital energy through drag very slowly until it sink to about half the stellar radius (from the surface) after which it will "plunge" down. They show that an exoplanet can not only increase the rotation of the host star (in a prograde orbit) but can, in some cases, reverse the rotation if in a retrograde orbit (see their Fig.~4 and 5). Given how dramatic  these effects are,  it is reasonable to expect that the turbulence (mixing length) will be  altered significantly.

While to estimate the effect precisely, 3D hydrodynamical simulations are needed, and this is beyond the scope of the present paper and therefore left for future work, we can do a rough estimate of the effect.  The relative energy deposited by the planet on the convective envelope is 
$\sim (M_{\rm p}/M_{\rm env}) (V_p/V_{env})^2$. This quantity is of order one. If all this energy goes in increasing the velocity of the convective blobs, this can represent a significant perturbation.  

 We should however consider that the energy of the planet will not be deposited  and absorbed by the stellar envelope instantly, but it will take about one dynamical time. The time between planet engulfment and when the star leaves the TRGB is  $\sim$  5 to  10   \% of the dynamical time. Hence only  5 to 10  \% of the orbital energy of the planet can go into the stellar envelope and affect the TRGB luminosity.
  
In mixing length theory, the velocity of the "blob" is proportional to the mixing length ($\alpha$) of this blob, so that any increase in velocity will increase the mixing length, provided the gradients of the environment and the "blob" are not modified significantly
\begin{equation}
v_{\rm b}^2 = g \delta (\nabla_b - \nabla_e) \frac{\alpha^2 g \rho}{8 P}
\end{equation}
where  $\delta(.)$ denotes  the change in its argument and the subscript $b$ refers to the "blob" while $e$ to the environment. So the the mixing length increase is expected to be of the same order. However, in detail, the  gradients of the environment and the "blob" will be modified. While this is fairly rough, if  the energy of the envelope changes by $5\sim10$ \% so will be the change in the  square of the velocity of the blobs, and therefore also of $\alpha^2$. As a consequence we estimate that  expected increase of the mixing length is at the 3~\% level. From Fig.~\ref{fig:models} this translates into 3\% changes in the luminosity of  the star at the TRGB.

It is clear that  detailed 3D hydro-dynamical simulations of the process of the engulfment of the gas giant by the star are needed  to asses accurately the increase in the turbulence on the convective layer and thus the effect on the luminosity of the TRGB. 

\section{Observationally quantifying mass-loss rate distributions  and frequency of engulfment}
We have found that  that  mass-loss rate and planets engulfment are the most prominent effects. One may worry that,   if unaccounted for, they might rend less robust the use of the TRGB as a distance indicator.  Therefore we consider whether it is possible to quantify these effects  on a galaxy-by-galaxy basis and therefore  account for them in the modeling. The dependence of the TRGB luminosity on galactic metallicity can be studied by comparing the observed values with those derived in 
Eq.\,\ref{eq:lum}. However, also the mass and the mixing length goes into Eq.\,\ref{eq:lum}. Since the TRGB luminosity is a narrow, well defined, function of the core mass, and since the horizontal branch morphology depends on the mass--to--core mass ratio, the mass and 
mass variation at the TRGB can be determined from the HB morphology, as was done in \cite{Jimenez96}. If the mass variation is 
ascribed to variations in $\eta$ (whatever cause this variation), then now also $\eta$ (i.e.\ the mass variation in Eq.\,\ref{eq:lum})
is determined. With a more strict quantification of the dependence of the mixing length parameter $\alpha$ on engulfment of planets into the host stars during RGB, then finally the quantitative dependence of Eq.\,\ref{eq:lum} on planetary morphology and mass loss
can be calculated and compared to observations. The value (and spread in value) of $\eta$ depends at least on the metallicity of the 
galactic environment the stars are made from, because the abundance of opacity--intensive dust grains that drives the mass loss is a direct function of metallicity. Since the abundance of gas giant exoplanets seems also to be a (complex) function of host star metallicity, the value of $\alpha$ will be a function at least of metallicity. Both of these parameters are therefore a function of the galactic environment in which the stars and their planetary systems are formed, and hence the TRGB luminosity is also a (complex) function of the galactic environment. We have sketched here the qualitative form and the magnitude of this function. The exact quantitative form of the function is not obvious, because of the complex non-linear form of the mutual dependence of $\eta$, $\alpha$, metallicity, and gas giant planetary abundance. On the other hand, even the qualitative considerations above predict that we should expect a large 
variation in the metallicity of observed RGB stars near the TRGB, and that this variation should be particularly large in low--metallicity systems. Galactic environments
where such a variation would not be found, will indicate that gas giant planets are not existing in terrestrial-like orbits
(born or migrated there) around stars in these environments, which will have severe implications on our knowledge about planetary system formation (and on the use of the TRGB method for cosmic distance determinations), and might contribute to the solution of the still mysterious fact that no giant exoplanets were found in the first HST transit experiment~\cite{Gilliland2000} and only few were found in the 
second HST transit survey~\cite{Sahu2006}. Hence, measurements of the variation in the metallicity of RGB stars can potentially impact two such different areas of research as the cosmological distance scale as well as the question about which environments allow formation of gas giant exoplanets. 

Planets are more frequent for higher metallicity environments. Ref.~\cite{Mortier} show that the cutoff is at $0.1$ the solar value. The planet formation rate is most likely not dependent on metallicity, but the giant planet formation rate is. This is because the formation of giant gas planets in the standard core accretion model requires that the core grows to $10$ Earth-masses before the gravitational force is big enough for a ($H/He$) run-away collapse transforming the solid core to a full gas giant planet. The growth to $10$ Earth-mass will usually only happen fast enough if there is sufficient abundance of $H_2 O$ (i.e. oxygen, hence metallicity) in the disk. The amount of heavy elements engulfed at the TRGB should therefore be independent on metallicity of the environment (galactic halo or disk), but the amount of momentum transfer into the TRGB star depends on metallicity and will be largest for high metallicity environments, for the reasons described above. The spectral effect at low metallicity should be markedly high and at high metallicity minimal, while the luminosity should correlate the opposite. This will provide us with the necessary correction  for improving the calibrations for RTGB as a distance indicator.

\section{Brown dwarf engulfment.}

Until now we have only considered the effect of giant planets on the TRGB luminosity. However, another class of objects that will have an important effect on modifying the TRGB luminosity is the engulfment of brown dwarfs.  Ref.~\cite{LivioBD} have studied this process in detail by modeling the accretion of brown dwarfs by around solar mass stars in the red giant phase. Note that the most likely phase for engulfment of the brown dwarf by the star is during the red giant phase. The brown dwarf will reach in some cases the core and detonate the He flash. It is illustrative to look at Fig.~1 in Ref.~\cite{LivioBD}. It is seen that for no accretion of a brown dwarf, the TRGB luminosity is independent of mass, as expected. However, when the brown dwarf is engulfed, the TRGB is terminated at lower luminosities. This will induce an additional bias if brown dwarf formation depends on the environment.

\section{Conclusions}
\label{sec:conclusions}

Because the TRGB distance method is  applied to galaxies that do not necessarily have the same chemical composition, or age, or environment of  the anchor galaxy (e.g., the Large Magellanic cloud), it is important to understand a possible environmental dependence of the TRGB. Here we have shown two important effects: star-to-star variations  in the mass-loss rate at the TRGB and the engulfment of part of the planetary system by the red giant star.  The variations in the mass-loss rate of the star at the TRGB can result in changes of up to 5\% of the star's the luminosity at the TRGB. On the other hand, the engulfment of gas giants by the red giant stars results in a systematic effect of decreasing the luminosity of the star at the TRGB by increasing the mixing length parameter. While we have shown that potentially the TRGB is more complex than initially thought, it will be necessary to investigate in detail what these environmental effects mean for the use of the TRGB luminosity as a distance indicator. We have also pointed out that these environmental effects can be mitigated by exploring the morphology of the horizontal branch (mass loss) and through high-resolution spectroscopy of stars at the red giant branch to  infer the frequency, and identify the engulfment, of exoplanets in low-metallicity red giant branch stars.

\acknowledgments

We thank the anonymous referee for many useful and constructive comments and specially for asking us to think about the effect of brown dwarf engulfment. We thank Jose Luis Bernal for useful comments on the draft. LV and RJ are supported by Spanish MINECO under project PGC2018-098866-B-I00 FEDER-EU. LV acknowledges support of European Union's Horizon 2020 research and innovation program ERC (BePreSySe, grant agreement 725327).

\end{document}